\documentclass[prd,twocolumn,showpacs,eqsecnum,amssymb,
amsmath,a4paper,superscriptaddress,nofootinbib]{revtex4}
\usepackage{graphicx,multirow}
\usepackage{epstopdf}
\usepackage{amsfonts}
\usepackage{mathrsfs}
\usepackage[innercaption]{sidecap}
\usepackage{overpic}
\def\d{{\mathrm{d}}}%
\begin{document}
%--------------------------------------------------------------
\setcounter{topnumber}{1}

\title{Sensitivity of Hawking radiation to superluminal dispersion relations}
%--------------------------------------------------------------
\author{C. Barcel\'{o}}
\affiliation{Instituto de Astrof\'{i}sica de Andaluc\'{i}a, CSIC, Camino Bajo de
Hu\'{e}tor 50, 18008 Granada, Spain}
%--------------------------------------------------------------
\author{L. J. Garay}
\affiliation{Departamento de F\'{i}sica Te\'{o}rica II, Universidad Complutense
de Madrid, 28040 Madrid, Spain}
\affiliation{Instituto de Estructura de la Materia, CSIC, Serrano 121, 28006
Madrid, Spain}
%--------------------------------------------------------------
\author{G. Jannes}
\affiliation{Instituto de Astrof\'{i}sica de Andaluc\'{i}a, CSIC, Camino Bajo de
Hu\'{e}tor 50, 18008 Granada, Spain}
\affiliation{Instituto de Estructura de la Materia, CSIC, Serrano 121, 28006
Madrid, Spain}
%--------------------------------------------------------------
%\author{S. Liberati}
%--------------------------------------------------------------

\date{\today}

%--------------------------------------------------------------
\begin{abstract}
%--------------------------------------------------------------

We analyze the Hawking radiation process due to collapsing
configurations in the presence of superluminal modifications of the
dispersion relation. With such superluminal dispersion relations, the
horizon effectively becomes a frequency-dependent concept. In
particular, at every moment of the collapse, there is a critical frequency
above which no horizon is experienced. We show that, as a consequence,
the late-time radiation suffers strong modifications, both quantitative
and qualitative, compared to the standard Hawking picture. Concretely,
we show that the radiation spectrum becomes dependent on the
measuring time, on the surface gravities associated with different
frequencies, and on the critical frequency. Even if the critical frequency
is well above the Planck scale, important modifications still show up.

%--------------------------------------------------------------
\end{abstract}
%--------------------------------------------------------------
\pacs{04.70.Dy, 04.62.+v, 04.60.Bc}
\maketitle

%--------------------------------------------------------------
\section{Introduction}
\label{S:intro}
%--------------------------------------------------------------
Phenomenological approaches to quantum gravity have recently started
to develop in parallel with the more traditional attempts to construct
such a theory from first principles. In particular, increasing
attention has focused on the consideration that maybe Lorentz
invariance is not a fundamental law, but an effective low-energy
symmetry which is broken at high energies (see,
e.g.,~\cite{Jacobson:2005bg} for a general introduction). Conceptually
speaking, in quantum gravity theories from first principles, it is not
really clear whether Lorentz invariance is fundamental or effective,
and in the latter case, how its breaking scale is related to the
Planck scale. For example, while many string theory scenarios
axiomatically incorporate Lorentz invariance, it has been argued that
in certain situations, violations of Lorentz invariance may occur in a
way consistent with world-sheet conformal
invariance~\cite{Mavromatos:2007xe}, thus leading to acceptable string
theory backgrounds. In the context of loop quantum gravity,
in~\cite{Gambini:1998it} it has been argued that quantum effects
should modify the relativistic dispersion relations, although the
issue seems far from settled (see~\cite{Ashtekar:2007px} for some
general remarks). In scenarios of emergent gravity
based on condensed matter
analogies~\cite{Barcelo:2005fc,Volovik:2003fe}, the situation is
clearer: Lorentz invariance is a low-energy effective symmetry, and so
it is expected to break at some scale, although not necessarily
related to (and therefore possibly much higher than) the Planck
scale~\cite{Klinkhamer:2005cp}.

A simple way of modelling a wide range of Lorentz violating effects
(and a quite natural one in the case of condensed matter analogies)
consists in modifying the dispersion relations at high
energy~\cite{Mattingly:2005re}. This modification can be subluminal or
superluminal, depending on whether high-frequency modes move slower or
faster than low-energy ones. We should mention that there also exist
ways of modifying the dispersion relations, for example by introducing
a minimum length or maximum energy, without violating Lorentz
invariance~\cite{Magueijo:2001cr}. Here, however, we will be
interested in modifications that explicitly break Lorentz
invariance. From extrapolation of current experiments, we know that
there exist stringent bounds on the most commonly expected types of
Lorentz violations at the Planck scale (see,
e.g.,~\cite{Jacobson:2005bg} and references therein). Nevertheless,
even violations at much higher energy scales might still significantly
affect black hole physics.

In this paper we will investigate the effects that a superluminal
modification of the dispersion relation would have on the Hawking
radiation produced by collapsing configurations. Hawking's original
prediction that black holes radiate
thermally~\cite{Hawking:1974rv,Hawking:1974sw}, rested on the implicit
assumption that the low-energy laws of physics, and in particular
Lorentz invariance, are preserved up to arbitrarily large scales, much
higher than the Planck energy. The question of the robustness of
Hawking's prediction to modifications of the transplanckian physics has
been tackled principally by analyzing effective field theories such that at
high energies a modification of the dispersion relation is 
incorporated (see however \cite{Agullo:2006um} for a different take on 
the problem).
Historically, attention has mainly been given to subluminal dispersion
relations. Important contributions such as
\cite{Unruh:1980cg,Jacobson:1991gr,Jacobson:1993hn,Unruh:1994je,Brout:1995wp,
Corley:1996ar,Corley:1996nw,Corley:1997pr}, and more recently
\cite{Unruh:2004zk,Schutzhold:2008tx} seem to have settled the robustness of
Hawking radiation with respect to subluminal modifications and truncation
of the frequency spectrum, although this conclusion still rests on certain
assumptions, usually related to the behaviour of the fields near the horizon. In any case, it is important to remember that this only solves part of the so-called `transplanckian problem' with regard to Hawking radiation. Indeed, subluminal modifications gradually dampen the influence of ultra-high frequencies, and so they do not explore arbitrarily large frequencies. So even assuming that it has been demonstrated that Hawking's result can be recovered without appeal to transplanckian frequencies, and that it is robust to subluminal modifications, the question remains whether it is also robust with respect to a (non-dampened) modification of the transplanckian physics. Superluminal modifications differ conceptually from subluminal ones, in that they gradually magnify the influence of ultra-high energies, and thereby offer an interesting test-case for the transplanckian robustness of Hawking radiation. 

An additional motivation for our study comes from a possible connection with experiment. It is well-known (see, e.g.,
\cite{Visser:2001kq}) that Hawking radiation is not a purely gravitational
effect, but a
characteristic of quantum field theory in curved spacetimes with a
horizon. Condensed matter systems such as Bose-Einstein condensates
(BECs) can, under certain approximations, be described by a
relativistic quantum field theory. It has also been shown, again under certain
idealizing
approximations, that (acoustic) black hole horizons in BECs are dynamically
stable~\cite{Barcelo:2006yi}\footnote{Note, however, that this dynamical stability is quite fragile. For example, a minimal amount of backscattering in the interior of the black hole could be sufficient to create dynamical instabilities~\cite{Barcelo:2006yi}, which in a realistic laboratory experiment might completely obfuscate any quantum radiation related to an analogue Hawking process. This instability was first revealed in a mode analysis of a black hole-white hole configuration under a superluminal dispersion relation~\cite{Corley:1998rk}. In this paper we will not worry about this issue and simply assume that the dynamics of the background is completely fixed.}. Therefore, BECs are expected to be good candidates for a possible experimental
verification
of Hawking radiation~\cite{Garay:2000jj,Garay:1999sk,Barcelo:2001ca,Barcelo:2000tg,Schutzhold:2006,Wuster:2007nf,Balbinot:2007de,Carusotto:2008ep,Wuester:2008kh}. But since the physics of BECs automatically leads to a superluminal dispersion relation at high energies, the
question is again which kind of modifications are to be expected in the laboratory realization of a BEC black hole with respect to the standard Hawking picture.

Of the above-mentioned works on the robustness of Hawking radiation, a
few have also tried to address superluminal modifications
\cite{Corley:1997pr,Unruh:2004zk,Schutzhold:2008tx}. Various problems make this
case quite different from the
subluminal one. These problems can be related to the fact that the
horizon becomes frequency-dependent when modifying the dispersion
relations. This is also true in the subluminal case, though
qualitatively in a very different way. With superluminal
modifications, the horizon lies ever closer to the singularity for
increasing frequencies, and asymptotically coincides with it. This causes the
`apparent' interior of the black hole (the interior of the zero-frequency horizon) to be exposed to the outside world. Since it
seems unreasonable
to impose a condition arbitrarily close to the singularity as long as
we do not have a solidly confirmed quantum theory of gravity, most of
the approaches used for subluminal dispersion are invalid, or at least
questionable, for superluminal dispersion. Moreover, since it seems
reasonable to expect that quantum effects will remove the general
relativistic singularity, a critical frequency might appear above which
no horizon would be experienced at all.

In the analysis that follows we will try to avoid making any
further assumptions about the physics near the singularity. For instance
we will analyze the characteristics of the radiation at retarded times at which
the singularity has not yet formed. Our approach will be based on a derivation of Hawking radiation through the
relation between the asymptotically past ray trajectories and the
asymptotically future ones in the case of a collapsing
configuration. The only assumption about these asymptotic extremities
are the standard ones, namely a Minkowski geometry in the asymptotic
past, and flatness at spatial infinity also in the asymptotic
future. The language we will use is related to the fluid analogy for
black holes, which provides the intuitive picture of the spacetime
vacuum flowing into the black hole and getting shredded in its center.

Our main results can be summarized as follows. Three crucial elements
distinguish the late-time radiation with superluminal dispersion
relations from standard relativistic ones. First, at any instant there
will be a critical frequency above which no horizon has yet been
experienced. This critical frequency will induce a finite limit in the
modes contributing to the radiation, which will therefore have a lower
intensity than in the standard case, even if the critical frequency is
well above the Planck scale. Second, due to the effective
frequency-dependence of the horizon, the surface gravity will also
become frequency-dependent and the radiation will depend on the
physics inside the black hole. Unless special conditions are imposed
on the profile to ensure that the surface gravity is nonetheless the
same for all frequencies below the critical one, the radiation
spectrum also undergoes a strong qualitative modification. Depending
on the relation between the critical frequency and the Planck
scale, the radiation from high frequencies is no longer negligible
compared to the low-frequency thermal part, but can even become
dominant. This effect becomes more important with increasing critical
frequency. Finally, a third effect is that the radiation will
extinguish as time advances.

The remainder of this article has the following structure. In section
\ref{S:model}, we will describe and motivate the classical geometry of our model and the concrete kinds of
profiles that we are interested in. In section \ref{S:SDR}, we will briefly review how
Hawking's standard result can be obtained in the formalism of our choice for the
case of standard relativistic dispersion relations. This calculation will be
adapted in section \ref{S:MDR} to superluminal dispersion relations, and we will
analytically obtain a formula for the late-time radiation for this case. Then,
in section \ref{S:graphics}, we will present graphics obtained from this
analytic formula by numerical integration. These graphics will illustrate the results mentioned in the previous paragraph, which we will discuss in more depth and compare with other results in the recent literature in section \ref{S:conclusion}.

%--------------------------------------------------------------
\section{The classical geometry}
\label{S:model}
%--------------------------------------------------------------

We will study radiation effects in simple 1+1 dimensional collapsing configurations.
It is well known that in the Hawking process, most of the radiation comes out
in the $s$-wave sector. Hence, a spherically symmetric treatment, effectively
1+1
dimensional, suffices to capture the most relevant aspects of the process.
We will work in Painlev\'e-Gullstrand coordinates, where
\begin{eqnarray}%
\d s^2=-[c^2-v^2(t,x)]\d t^2 - 2v(t,x)\d t \d x +\d x^2~,
\label{acoustic-metric}
\end{eqnarray}%
and make the further simplification of taking a constant $c$ (we will
always write $c$ explicitly to differentiate it from the
frequency/wave-number dependent $c_k$ that will show up in the
presence of a superluminal dispersion relation). In this manner all
the information about the configuration is encoded in the single
function $v(t,x)$. The Painlev\'e-Gullstrand coordinates have the
advantage, compared to the Schwarzschild form, of being regular at the
horizon. Moreover, they suggest a natural interpretation in terms of
the language of acoustic models. In such a model, $c$ represents the
speed of sound and $v$ is the velocity of the fluid (which corresponds
to the velocity of an observer free-falling into the black hole).
Throughout the paper we will consider the fluid to be left-moving, $v
\leq 0$, so that the outgoing particles of light/sound move towards
the right.

\begin{figure}
\begin{center}
\includegraphics[width=0.9\columnwidth]{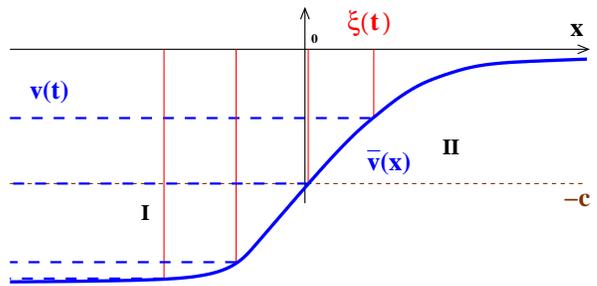}
\end{center}
\bigskip
\caption{\label{Fig:v-constant-kappa} Velocity profile of a black hole with a linear slope of the velocity $\bar{v}(x)$, and hence a constant surface gravity $\kappa_{\omega'}$, from the classical horizon at $x=0$ down to some predefined limiting value. The auxiliary function $\xi(t)$ separates each instantaneous velocity profile $v(t,x)$ into a dynamical \mbox{region I}: $v(t,x)=v(t)=\bar{v}(\xi(t))$ and a stationary region II: $v(t,x)=\bar{v}(x)$.}
\end{figure}

\begin{figure}
\begin{center}
\includegraphics[width=0.9\columnwidth]{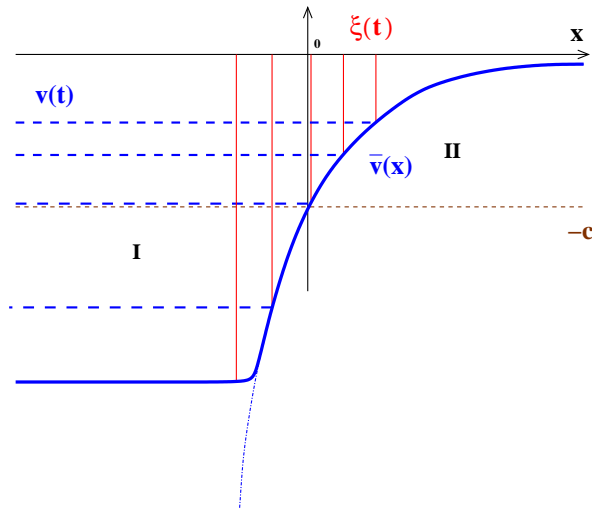}
\end{center}
\bigskip
\caption{\label{Fig:v-increasing-kappa} Velocity profile characteristic
of a Schwarzschild black hole. The slope of $\bar{v}(x)$ increases from the classical horizon at $x=0$ leftwards up to a constant limiting value, leading to a surface gravity $\kappa_{\omega'}$ which increases with the frequency.}
\end{figure}

The most relevant aspects of the analysis presented hereafter
depend only on the qualitative features of the fluid profile. However, to 
justify some specific calculations later on it is helpful to use the following 
concrete profiles, see figs.~\ref{Fig:v-constant-kappa} and
\ref{Fig:v-increasing-kappa}. Consider a velocity profile $\bar{v}(x)$
such that \mbox{$\bar{v}(x\to +\infty)=0$}, \mbox{$\bar{v}(x=0)=-c$} and further
decreasing monotonically as \mbox{$x\to -\infty$}. In the fluid image, the
fluid nearly stands still at large distances and accelerates inwards,
with a sonic point or horizon at $x=0$. The further decrease in
$\bar{v}$ can either be linear until a constant limiting value is almost
achieved, as in
fig.~\ref{Fig:v-constant-kappa}, or of the form 
\begin{eqnarray}%
\bar v(x)=-c\sqrt{2M/c^2 \over x +2M/c^2}~,
%\label{acoustic-metric}
\end{eqnarray}%
again up to a constant limiting value, as in
fig.~\ref{Fig:v-increasing-kappa}. Let us recall that this second
velocity profile corresponds to the Schwarzschild line element with $M$ 
the central mass, as can be seen~\cite{Visser:1997ux} by reparametrizing the time coordinate and using $r=x+2M/c^2$ as radial coordinate.

%====================================================
%====================================================

%====================================================
%====================================================

Up to here, we have described a stationary profile. Now, to incorporate
the dynamics of the collapse, let us introduce a
monotonically decreasing function $\xi(t)$, \mbox{$\xi(t \to -\infty) \to
+\infty$}, and define $v(t,x)$ as
\begin{eqnarray}\label{profile}
v(t,x)~ = 
\left \{ 
\begin{array}{lll}
\bar{v}(\xi(t)) &~\text{if}~ &x \leq \xi(t)~,\\
\bar{v}(x) &~\text{if}~ &x \geq \xi(t)~.
        \end{array}
\right.
\end{eqnarray}
Imagining the collapse of a homogeneous star, the function $\xi(t)$
represents the distance from the star surface to its gravitational (Schwarzschild) radius. The dynamical configuration that
we obtain
consists of a series of snapshots. In each snapshot, $|v(t,x)|$
increases (i.e., the fluid accelerates) from $\bar{v}=0$ for $x\to
+\infty$, up to a point $x_0=\xi(t)$, and then remains constant as $x$
further decreases towards $-\infty$. For consecutive snapshots, the
point $x_0=\xi(t)$ moves leftwards, so $v(t,x)$ covers an ever
larger part of $\bar{v}(x)$.

From the point of view of an outgoing (right-moving) particle of
light/sound, the configuration is nicely split up into, first, a
dynamical or $t$-dependent region, and then a stationary $x$-dependent
region, as defined in \eqref{profile}.\footnote{In a physically
realistic model, the apparent kink in the profile where the transition
between both regions takes place will of course be smoothed out.}

The difference between the two types of profiles,
figs.~\ref{Fig:v-constant-kappa} and \ref{Fig:v-increasing-kappa}, can
best be explained in terms of the surface gravity. When modifying the
dispersion relation, the horizon becomes a frequency-dependent
concept: each frequency experiences a different horizon, 
as we will see explicitly in section~\ref{S:MDR}. In
particular, for superluminal modifications, the horizon forms later
(i.e., at higher values of $|\bar{v}|$, or more negative values of
$x$) for increasing frequencies. The surface gravity will then also
become frequency-dependent:
\begin{eqnarray}\label{def-kappa}
\kappa_{\omega'}\equiv c \bigg|{\d \bar v \over \d x} \bigg|_{x=x_{H,\omega'}}~,
\end{eqnarray}
where $x_{H,\omega'}$ is the frequency-dependent location of the
horizon.  This surface gravity will be seen in section~\ref{S:MDR} and
\ref{S:graphics} to play a crucial role. This explains our choice of 
the two types of profiles: In our first profile,
fig.~\ref{Fig:v-constant-kappa}, we have considered a linear velocity
profile $|\bar{v}(x)|$ from the classical or zero-frequency horizon
($c=|v|$) down to some given limiting value, $|v|_{\rm max}$, from
where it stays constant. This maximum velocity defines a critical
frequency $\omega'_c$ as we will discuss in more detail in
section~\ref{S:MDR}. Then the surface gravity $\kappa_{\omega'}$ will
be frequency-independent up to this critical frequency $\omega'_c$ after
which it will rapidly vanish. In the second profile, see
fig.~\ref{Fig:v-increasing-kappa}, we have taken a velocity profile
typical for a Schwarzschild black hole, and therefore
$\kappa_{\omega'}$ increases with $\omega'$, again up to a predefined
limiting value corresponding to the horizon for a critical frequency
$\omega'_c$. In the context of the fluid analogy, it seems obvious
that some mechanism will avoid the formation of a singularity. But in
any case, we will also take into account the limit for an infinite
critical frequency, which corresponds to a velocity profile with 
$|v|_{\rm max} \to \infty$.

%--------------------------------------------------------------
\section{Standard Dispersion Relations}
\label{S:SDR}
%--------------------------------------------------------------

In this section we will briefly review a way of deriving Hawking's
formula for the radiation of a black hole with standard (relativistic)
dispersion relations of the form \mbox{$\omega^2=c^2k^2$}. 
For the sake of simplicity, we will only consider a massless scalar field.
We will summarize the main steps: from the Klein-Gordon inner product and the
definition of the Bogoliubov $\beta$ coefficients, over the relation
between past and future null coordinates, to the black-body radiation
in the wave-packet formulation. Our aim is to present the key points of the procedure in
such a way that they can easily be adapted to superluminally
modified dispersion relations---the subject of the next sections.

\subsection{Inner product}
%-------------------------------------------------------------------
The d'Alembertian or wave equation for a massless scalar field in 1+1
dimensions for the metric~(\ref{acoustic-metric}) with constant $c$
can be written as
\begin{eqnarray}%
&&(\partial_t+\partial_x v)(\partial_t+v\partial_x)\phi = c^2 \partial_x^2
\phi~.
\end{eqnarray}
This conformally invariant theory is equivalent to the
dimensionally-reduced 3+1 spherically-symmetric theory if one neglects
the backscattering due to the angular-momentum potential barrier
(responsible for the so-called grey-body factors).

In the space of solutions of this wave equation, we can define the Klein-Gordon
pseudo-scalar product
\begin{eqnarray}%
(\varphi_1,\varphi_2) \equiv 
-i \int_\Sigma \d\Sigma^\mu ~ 
\varphi_1 \stackrel{\leftrightarrow}{\partial_\mu}\varphi_2^*~,
\end{eqnarray}%
which is independent of the choice of the spatial slice $\Sigma$. For a
\mbox{$t=$constant} slice, and in particular for \mbox{$t\to +\infty$,} this
becomes
\begin{equation}\label{varphi1-varphi2}%
(\varphi_1,\varphi_2) =
-i \int \d x ~ 
\left[\varphi_1 (\partial_t+v\partial_x )\varphi_2^* 
-\varphi_2^* (\partial_t+v\partial_x )\varphi_1\right]~.
\end{equation}%
We can define future null coordinates $u(t,x)$ and $w(t,x)$, such that, when \mbox{$t,x\to +\infty$},
\begin{eqnarray}%
u(t,x) \to t - x/c~,&&~~
w(t,x) \to t + x/c~.
\end{eqnarray}%

%\begin{eqnarray}%
%\lim_{t,x\to +\infty } u(t,x) &=& t - x/c~,\\
%\lim_{t,x\to +\infty } w(t,x) &=& t + x/c~.
%\end{eqnarray}%
%
Writing the inner product in terms of these null coordinates, gives, for the
limit $t \to +\infty$,
\begin{eqnarray}\label{scalar_product}%
(\varphi_1,\varphi_2) = &&\hspace{-4mm}
-{i c \over 2}\left\{
\int_{-\infty}^{+\infty} \d u  ~ 
\left[\varphi_1 \partial_u\varphi_2^* 
-\varphi_2^* \partial_u\varphi_1\right]_{w=+\infty}
\right.
\nonumber
\\
&&\hspace{-10mm}
\left.
+\int_{-\infty}^{+\infty} \d w ~ 
\left[\varphi_1 \partial_w\varphi_2^* 
-\varphi_2^* \partial_w\varphi_1\right]_{u=+\infty}
\right\}~.
\label{KG-product}
\end{eqnarray}%
Note that, in the derivation of this formula, we have only
made use of coordinate transformations, without any appeal to their
null character or to geometrical tools such as conformal diagrams or
the deformation of Cauchy surfaces. This is important because in
the case of modified dispersion relations, such geometrical concepts
become problematic and actually can only be maintained in the context
of a rainbow geometry---if at all.

Similarly we could have used past null coordinates $(U,W)$, which obey, when $t\to -\infty$,
\begin{eqnarray}%
U(t,x) \to t - x/c~,&&~~
W(t,x) \to t + x/c~,
\end{eqnarray}%
%
%\begin{eqnarray}%
%\lim_{t\to -\infty } U(t,x) = t - x/c~,\\
%\lim_{t\to -\infty } W(t,x) = t + x/c~,
%\end{eqnarray}%
%
to calculate the inner product in the asymptotic past.

\subsection{Bogoliubov $\beta$ coefficients}
%-------------------------------------------------------------------

The right-moving positive-energy solutions associated with the 
asymptotic past and with the asymptotic future, normalized in the Dirac-delta sense, can be expressed as 
\begin{align}\label{modes-past+future}%
\psi'_{\omega'}={1 \over \sqrt{2\pi c ~ \omega'}} e^{-i\omega' U}~,&&
\psi_{\omega}={1 \over \sqrt{2\pi c ~ \omega}} e^{-i\omega u}~,
\end{align}
respectively, where we use primes to indicate asymptotic past values.

The relevant Bogoliubov $\beta$ coefficients encoding the production of 
radiation are defined as \mbox{$\beta_{\omega \omega'} \equiv (\psi'_{\omega'},\psi_{\omega}^*)$}.
%
%\begin{eqnarray}\label{def-beta}%
%\beta_{\omega \omega'} \equiv (\psi'_{\omega'},\psi_{\omega}^*)~.
%\end{eqnarray}%
%
The mode mixing relevant for the Hawking process occurs in the 
right-moving sector. Therefore, we only need to calculate 
the first term in the scalar product (\ref{KG-product}).
Then, plugging \eqref{modes-past+future} into the definition of $\beta$ and 
integrating by parts we obtain the simple expression
\begin{eqnarray}\label{beta}%
\beta_{\omega \omega'} =
{1 \over 2 \pi} \sqrt{\omega \over \omega'}
\int \d u  ~  e^{-i \omega' U(u)} e^{-i \omega u}~,
\end{eqnarray}%
so that all the information about the produced radiation is contained in the 
relation $U=U(u) \equiv U(u,w\to +\infty)$.

\subsection{Relation $U(u)$}
\label{SS:relation_U(u)}
%-------------------------------------------------------------------

For a standard relativistic dispersion relation, it is well known that 
the relation between $U$ and $u$ for a configuration that forms a horizon 
(in our case at $x=0$) can be expressed at late times as \mbox{$U=U_H-Ae^{-\kappa u/c}~$} (note that we use a subscript $H$ for all quantities associated with the horizon), where $U_H$, $A$ and the surface gravity 
\begin{eqnarray}\label{standard-kappa}
\kappa \equiv c\bigg| \frac{\d \bar{v}}{\d x}\bigg|_{x=0}
\end{eqnarray}
are constants.

We can define a threshold time $u_I$ at which an asymptotic
observer will start to detect thermal radiation from the black hole.
This retarded time corresponds to the moment at which the function
$U(u)$ enters the exponential regime.
We can then rewrite the previous expression, valid for $u>u_I$, as
\begin{eqnarray}%
U=U_H-A_0 e^{-\kappa (u-u_I)/c}~.
\end{eqnarray}
Plugging this relation into \eqref{beta} and integrating in $u$ gives
\begin{eqnarray}\label{U-u_step1}%
\beta_{\omega\omega'} &=
&{1 \over 2\pi }\sqrt{\omega \over \omega'} \;
{c \over \kappa}\exp[- i\omega' U_H] 
\exp[-i\frac{c\omega}{\kappa}\ln(\omega'A_0)]\nonumber\\
&&~~\times \exp(-i\omega u_I)\exp(-{\pi c\omega \over 2 \kappa})
\Gamma(ic\omega/\kappa )~.
\end{eqnarray}%

\subsection{Wave packet formulation}
\label{SS:wave-standard}
%-------------------------------------------------------------------

In order to obtain physically sensible results, it is a good
precautionary measure to replace the monochromatic rays described
until now by wave packets (see e.g.~\cite{Hawking:1974sw} or the
discussion in~\cite{Fabbri:2005mw}). Positive-energy wave packets can be
defined as
\begin{eqnarray}
P_{\omega_j,u_l}(\omega) \equiv \bigg\{ 
\begin{tabular}{ll}
${e^{i\omega u_l} \over \sqrt{\Delta \omega}}$ &~
$- {1 \over 2}\Delta \omega  <\omega -\omega_j< {1 \over 2}\Delta
\omega $\\
$0$ & ~~{\rm otherwise},
\end{tabular}
\end{eqnarray}
where $u_l \equiv u_0+2\pi l/\Delta \omega$ with $u_0$ an overall reference and
$l$ an integer phase parameter. The central
frequencies of the wave 
packets are \mbox{$\omega_j \equiv j\Delta \omega$}, with $\Delta \omega$ their width.

Then, from the expression \mbox{$\beta_{\omega_{j},u_{l};\omega'} \equiv
\int \d\omega \;
\beta_{\omega\omega'}
P_{\omega_j,u_l}(\omega)~,$} 
and assuming that the wave packets are sufficiently narrow
($\Delta\omega\ll\omega_j$), we obtain
\begin{eqnarray}\label{U-u_step2}
|\beta_{\omega_{j},u_{l};\omega'}|^2 
\approx \frac{c\Delta \omega}{2\pi\omega'\kappa}\frac{\sin^2(z-z_l)}{(z-z_l)^2} 
{1 \over \exp({2\pi c\omega_j \over \kappa})-1}~,
\end{eqnarray}
where we have defined
\begin{align}
z=\frac{c\Delta\omega}{2\kappa}\ln\omega' A_0~,&&~~
z_l=\frac{\Delta\omega}{2}(u_l-u_I)~.
\end{align}
Finally, integration in $\omega'$ gives the number of particles with frequency
$\omega_j$ detected at time $u_l$ by an asymptotic observer:
\begin{eqnarray}\label{N-standard}%
N_{\omega_j,u_l}& = &\int_0^{+\infty}\d\omega'~
|\beta_{\omega_j,u_l;\omega'}|^2~
\\
& \approx &{1 \over \exp(2\pi c\omega_j/\kappa) -1}~,
\end{eqnarray}%
which reproduces Hawking's formula (in the absence of backscattering) and 
corresponds to a Planckian spectrum with temperature $T_H=\kappa/(2\pi c)$.

%--------------------------------------------------------------
\section{Superluminally modified dispersion relations}
\label{S:MDR}
%--------------------------------------------------------------

In this section we will indicate how the late-time radiation
originating from the formation of a black hole from collapse can be
calculated in the case of superluminal dispersion relations.

We introduce superluminally modified dispersion relations by adding a
quartic term to the wave equation:
\begin{eqnarray}\label{modified_laplace}%
(\partial_t+\partial_x v)(\partial_t+v\partial_x)\phi = 
c^2 \left(\partial_x^2 +{1 \over k_P^2}\partial_x^4 \right)\phi~,
\end{eqnarray}
where $k_P$ (the `Planck scale') is the scale at which non-relativistic deviations in the associated dispersion relation
\begin{eqnarray}%
(\omega - vk)^2= c^2k^2\left(1+\frac{k^2}{k_P^2}\right)~
\label{dispersion}
\end{eqnarray}%
become significant. We use this relation for concreteness and because it is the one
that shows up in Bose--Einstein condensates, where $k_P=2/\xi$, with
$\xi$ the healing length of the condensate. However, qualitatively,
our results will not depend on the specific form of the deviations from the
relativistic dispersion relation but on their
superluminal character.

This dispersion relation leads to a modification in both the phase velocity $v_{ph}$ and the group velocity $v_g$. 
%
%====================================================
\begin{figure}
\begin{center}
\includegraphics[width=0.9\columnwidth]{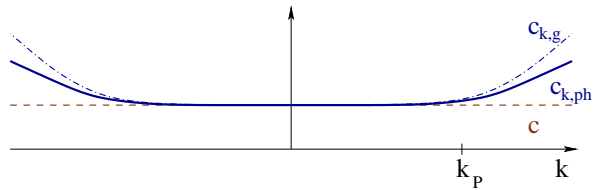}
\end{center}
\bigskip
\caption{\label{Fig:c_k} Behaviour of the effective 
phase $c_{k,ph}$ and group $c_{k,g}$ speeds of light/sound with respect to the wave number $k$. Due to the non-relativistic dispersion relation, the effective velocities become `superluminal' for $k>k_P$ (where $k_P$ is the Planck scale).
}
\end{figure}
%====================================================
%
For a right-moving wave, we have
\begin{align}
v_{ph} \equiv \frac{\omega}{k} =c_{k,ph}+v~, &&
v_g\equiv \frac{\d \omega}{\d k} =c_{k,g}+v~,
\end{align}
where we have introduced the effective $k$-dependent phase and group speeds of light/sound
\begin{align}\label{group_speeds}
c_{k,ph}=c\sqrt{1+\frac{k^2}{k_P^2}}~, &&
c_{k,g}=c\frac{1+2\frac{k^2}{k_P^2}}{\sqrt{1+\frac{k^2}{k_P^2}}}~,
\end{align}
respectively (see fig.~\ref{Fig:c_k}). Both $c_{k,ph}$ and $c_{k,g}$ become larger than $c$ (i.e., become `superluminal') as $k$ increases above $k_P$.

At first sight, it seems obvious that the
ray equation should be defined in terms of the group velocity
$v_g$. Nevertheless, the question of whether the velocity relevant for
Hawking radiation is the group or the phase velocity seems to be
tricky~\cite{Schutzhold:2008tx,Visser:2007du,Visser:2007nx}. For example,
\eqref{stationary} below suggests that the phase velocity might be
relevant. We limit ourselves to remark that $c_{k,g}$ and $c_{k,ph}$
show the same qualitative behaviour. Hence our results are independent
of this question and we will simply write $c_k$ (or $c_{k(\omega')}$
when wishing to emphasize the frequency-dependence). Then,
there will be a horizon, which becomes frequency-dependent, when $c_k+v=0$,
irrespectively of whether $c_{k,g}$ or $c_{k,ph}$ is used for $c_k$.
Moreover, since $c_k$ becomes arbitrarily high for increasing wave
number $k$, given a certain $|v|_{\rm max}$ at a particular instant of
time, there will be a critical $\omega'_c$ such that waves with an
initial frequency $\omega'>\omega'_c$ do not experience a horizon at
all. The only exception to this rule occurs when the velocity profile
ends in a singularity $\bar{v}\to -\infty$, which implies $\omega'_c
\to \infty$.

Our aim is to calculate the black hole radiation with superluminally
modified dispersion relations. We will now repeat the main steps of
section~\ref{S:SDR}, and point out where and how these modifications
must be taken into account.

\subsection{Generalization of inner product}
\label{SS:generalization_inner_product}
%-------------------------------------------------------------------

The essential point with regard to the pseudo-scalar
product \eqref{varphi1-varphi2} is the following. Its explicit form
for $t=$constant is not changed by the presence of the $\partial_x^4$
term in the wave equation (see also the discussion in~\cite{Corley:1998rk}). Indeed, it is still a well-defined inner
product, and in particular it is still a conserved quantity, since
\begin{eqnarray}
\partial_t(\varphi_1,\varphi_2)=\int \d x~
\varphi_1\stackrel{\leftrightarrow}{\partial^4_x}\varphi_2^*~,
\end{eqnarray}
which can be seen, by repeated integration by parts, to vanish under the usual
assumption that the fields die off asymptotically. Note that the
modification of the dispersion relation singles out a preferred time
frame: the `laboratory' time $t$. Changing to another time $\tilde t$
will in general lead to a mixing between $t$ and $x$, and hence the
simple relations given here would no longer be valid.

Using the preferred time $t$, and making exactly the same change of
coordinates as in the case of standard dispersion relations, we can
again transform the inner product, evaluated at $t \to +\infty$, into
the expression \eqref{scalar_product} in terms of $u$ and $w$. 
As in the standard case, only the first term is relevant for the Hawking process:
\begin{eqnarray}\label{scalar_U}%
-{i c \over 2}
\int_{-\infty}^{+\infty} \d u  ~ 
\left[\varphi_1 \partial_u\varphi_2^* 
-\varphi_2^* \partial_u\varphi_1\right]_{w=+\infty}~.
\end{eqnarray}

Note that we are now using $u$ and $w$ merely as a perfectly good set of
(auxiliary) coordinates, in order to cast the inner product into a
useful form. However, we cannot associate them any null character such as 
they had in the case of standard dispersion relations.

\subsection{Rainbow null coordinates}
\label{SS:Rainbow}
%-------------------------------------------------------------------

Let us define some sets of spacetime functions that will prove
to be useful in what follows. Given a fluid profile of the type 
described in section~\ref{S:model}, we can integrate the ray equation 
\begin{eqnarray}%
{\d x \over \d t}=c_{k(\omega')}(t,x)+v(t,x)~,
\end{eqnarray}
starting from the past left infinity towards the right. The
ray initially has a frequency $\omega'$ and an associated initial
wave number $k'$, from which we can deduce the value of $c_{k'}$. In
the left region, where the velocity profile is dynamic but
position-independent ($v(x,t)=v(\xi(t))$, with $\xi(t)$ the auxiliary function introduced in section~\ref{S:model}), $k=k'$ can be considered as fixed
while the frequency changes (this is what happens in a mode
solution of equation~\eqref{modified_laplace} in this region). Then,
we can define the function
\begin{eqnarray}%
{\cal U}_{\omega'}(t,x)= 
{1 \over \omega'}\int {\bar \omega}(t)\d t -{k' \over \omega'} x~,
\end{eqnarray}
where ${\bar \omega}(t)$ is the instantaneous frequency of the
particle at each time $t$, defined through the dispersion
relation and such that \mbox{${\bar \omega}(t \to -\infty)=\omega'$}.

When the ray reaches the kink it passes into a stationary region in
which the velocity profile only depends on the position ($v(t,x)=\bar v(x)$). At
the kink, the ray still has the initial wave number $k'$ and
a frequency $\omega$. In its propagation towards the right this
frequency now remains fixed while its wave number becomes a decreasing
function $\bar k (x)$ of the position, such that $\bar k (\xi(t_0))=k'$, with $t_0$ the moment at which the kink is crossed. The final frequency of the ray will then simply be $\omega$ and its final wave number $k=
\lim_{x \to +\infty} \bar k(x)$. In this region, then,
the function ${\cal U}_{\omega'}(t,x)$ can be expressed as 
\begin{eqnarray}\label{stationary}%
{\cal U}_{\omega'}(t,x) = t - \int {{\bar k}(x) \over \omega} \d x~.
\end{eqnarray}
Note here that $\omega$ and $\bar k(x)$ both depend on the initial $\omega'$.

The same can be done by integrating the ray equation starting from the
future. In this way we can define $u_\omega$ functions. (The same procedure can be used to define ${\cal W}_{\omega'}$ and $w_{\omega}$). It is worth noting that ${\cal U}_{\omega'}$ and $u_\omega$ are not null coordinates in the usual geometric sense, since they are frequency-dependent, but they nevertheless share many properties with null coordinates.

\subsection{Bogoliubov $\beta$ coefficient}
%-------------------------------------------------------------------

The calculation of the inner product (\ref{scalar_U}) involves
the limit $w \to +\infty$ (equivalently, $w_\omega \to +\infty$). 
So, the general form 
\begin{eqnarray}%
u_{\omega}= u_{\omega}(u,w)
\end{eqnarray}
can be simplified to $u_{\omega}= u_{\omega}(u)\equiv u_\omega(u,w\to +\infty)$. We can therefore
change variables in the inner product from $u$ to $u_{\omega}$. The
combination of the derivative and the integral means that the form of
\eqref{scalar_U} is preserved in the new integration variable $u_{\omega}$.
Then we can write the Bogoliubov $\beta$ coefficients relevant for the Hawking process as
\begin{equation}
\beta_{\omega \omega'}=
-{i c \over 2}
\int_{-\infty}^{+\infty} \d u_{\omega}  
[\psi'_{\omega'} \partial_{u_{\omega}}\psi_{\omega} 
-\psi_{\omega} \partial_{u_{\omega}}\psi'_{\omega'}
]_{w_{\omega}=+\infty}~.
\label{KG-modified-product}
\end{equation}

Now, assuming that the profiles vary slowly (in scales much larger than the
Planck distance), the right-moving positive-energy modes associated with
past and future infinity can be approximated by the following simple
expressions:
\begin{eqnarray}\label{modes_MDR_P}%
\psi'_{\omega'} \approx {1 \over \sqrt{2\pi~c~\omega'}} e^{-i\omega' ~{\cal
U}_{\omega'}}~,
\end{eqnarray}
\begin{eqnarray}\label{modes_MDR_F}%
\psi_{\omega} \approx {1 \over \sqrt{2\pi~c~\omega}} e^{-i\omega ~u_{\omega}}~,
\end{eqnarray}
so that the Bogoliubov coefficients read
\begin{eqnarray}\label{beta_MDR}%
\beta_{\omega \omega'} \approx
{1 \over 2 \pi} \sqrt{\omega \over \omega'}
\int \d u_{\omega}  ~  
e^{-i \omega' \mathcal{U}_{\omega'}(u_{\omega})} e^{-i \omega u_{\omega}}~.
\end{eqnarray}%

In analogy with the standard case, all information about the radiation
is seen to be encoded in the relation $\mathcal{U}_{\omega'}(u_{\omega})$. In
this expression $\omega'$ is the initial frequency of a ray at the past
left infinity and $\omega=\omega(\omega')$ its final frequency when
reaching the future right infinity.

The previous approximation for the Bogoliubov coefficients amounts to
considering profiles that vary slowly both with $x$ and $t$. This is
equivalent to considering large black holes. In general the quartic
term in the wave equation, or equivalently the quartic modification of
the dispersion relation, introduce a new source of backscattering, on
top of the usual angular-momentum potential barrier which we have
already neglected. In our approximation, this additional
backscattering (beyond the standard grey-body factors) has been
neglected. For large black holes this contribution will in any case be
very small as has been observed in numerical simulations~\cite{Schutzhold:2008tx}. In addition we are also neglecting any reflection caused
by the kink. However, let us remark that given an approximative scheme
for calculating $\mathcal{U}_{\omega'}(u_{\omega})$ for general
profiles $v(t,x)$, the same $\mathcal{U}_{\omega'}(u_{\omega})$
obtained from a profile with a kink would be obtainable from one (or
several) specific $v(t,x)$, this time perfectly smooth and thus
causing no further backscattering. Our results, which rely only on the
specific form of the relation $\mathcal{U}_{\omega'}(u_{\omega})$, are
therefore valid beyond the specific configurations with a kink
presented in this paper.

\subsection{Relation $\mathcal{U}_{\omega'}(u_{\omega})$}
%-------------------------------------------------------------------
Our next task is to calculate the uniparametric family of functions $\mathcal{U}_{\omega'}$ and the relation between $\mathcal{U}_{\omega'}$ and $u_{\omega}$ for different configurations. As explained in section~\ref{SS:Rainbow}, the relation $\mathcal{U}_{\omega'}(u_{\omega})$ is obtained by integrating the
ray equation \mbox{$\d x/\d t=c_k+v$} using the profiles discussed in
section~\ref{S:model}. These can be described by means of
stationary profiles $\bar{v}(x)$ and an auxiliary function $\xi(t)$,
see the definition of $v(t,x)$ in \eqref{profile}. We will use a straightforward extension of the procedure established in~\cite{Barcelo:2006np} (see also~\cite{Barcelo:2006uw} for a summary) for a relativistic dispersion relation. Care must be taken, however, with the quantities that depend on the frequency $\omega'$. In particular:
\begin{itemize}
\item We will denote by $x_{H,\omega'}$ and $t_{H,\omega'}$ the 
position and the time at which the horizon associated with a particular 
initial frequency $\omega'$ is formed.
\item The surface gravity $\kappa_{\omega'}$, defined in
\eqref{def-kappa}, allows to write, for all $\omega'<\omega'_c$ and for $x$
close to $x_{H,\omega'}$:
 \begin{eqnarray}
 \bar{v}(x) \approx -c_k+\frac{1}{c}\kappa_{\omega'}(x-x_{H,\omega'})~,
 \end{eqnarray}
up to higher-order terms in $x-x_{H,\omega'}$.
\item The linearization of $\xi(t)$ near $t_{H,\omega'}$ also requires the
introduction of an $\omega'$-dependent parameter $\lambda_{\omega'}$:
 \begin{eqnarray}\label{xi}
 \xi(t)  \approx x_{H,\omega'}-\lambda_{\omega'}(t-t_{H,\omega'})~,
 \end{eqnarray}
again up to higher-order terms.
\end{itemize}
Let us consider laboratory times $t>t_{H,0}\equiv t_{H,\omega'=0}$ such that $\xi(t)$
has already crossed the classical horizon at \mbox{$x=0$}: \mbox{$\xi(t>t_{H,0})<0$}.
As we explained earlier we can define a critical frequency $\omega_c'$ as the
minimum initial frequency such that $c_k+\bar
v(\xi(t))> 0$ for all $\omega'>\omega_c'$, or, in other words, the minimum
frequency which at that particular time has not experienced any horizon yet.

In the dynamical part of the profile, where $v=\bar{v}(\xi(t))$, for rays crossing the kink just before the formation of the horizon, integration of the ray equation leads to
\begin{eqnarray}\label{U_vs_U_H}
{\cal U}_{\omega'}
\approx {\cal U}_{H,\omega'} +
\frac{\lambda_{\omega'}}{c_{k'}}(t_{0,\omega'}-t_{H,\omega'})~,
\end{eqnarray}
for small values of $|t_{0,\omega'}-t_{H,\omega'}|$ (see~\cite{Barcelo:2006np} for details of the calculation in the case of a relativistic dispersion relation), where $t_{0,\omega'}$ is the (frequency-dependent) time at which the kink separating the dynamical and the stationary regions is crossed, \mbox{$c_{k'}=\lim_{t\to
-\infty}c_{k(\omega')}$} is the speed of light in the asymptotic past, and
\begin{equation}
{\cal U}_{H,\omega'} \equiv t_{H,\omega'} -\frac{\xi(t_{H,\omega'})}{c_{k'}}+
\frac{1}{c_{k'}}\int_{-\infty}^{t_{H,\omega'}}\bar{v}(\xi(t)) \d t~
\end{equation}
is the ray constituting the horizon associated with the frequency $\omega'$.

In the stationary part of the profile, where $v=\bar{v}(x)$, we obtain, 
for \mbox{$t_{0,\omega'}\to t_{H,\omega'}$}, again using a method and notation based on~\cite{Barcelo:2006np},
\begin{equation}
u_{\omega} \approx t_{0,\omega'}-\frac{\xi(t_{0,\omega'})}{c_{k,f}}+C_1
-\frac{c}{\kappa_{\omega'}}\ln[\xi(t_{0,\omega'})-x_{H,\omega'}]~,
\end{equation}
where $c_{k,f}=\lim_{t\to +\infty}c_{k(\omega')}$ is the speed of light in the asymptotic future, and $C_1$ a bulk integration constant. So
\begin{eqnarray}\label{u_vs_xi}
\xi(t_{0,\omega'})-x_{H,\omega'} = C_{\omega'}e^{-\kappa_{\omega'}u_{\omega}/c}.
\end{eqnarray}

Both regions are connected by combining \eqref{U_vs_U_H} and \eqref{u_vs_xi}.
Making use of \eqref{xi}, this gives
\begin{eqnarray}\label{U-u_MDR}
{\cal U}_{\omega'}&=&{\cal U}_{H,\omega'} - A_{\omega'} e^{-\kappa_{\omega'}
u_{\omega}/c}~.
\end{eqnarray}
This relation is valid for frequencies for which a horizon is
experienced, i.e.\ for $\omega'<\omega'_c$ and times $u_{\omega}>u_{I,\omega'}$, where $u_{I,\omega'}$ is the
threshold time defined in section~\ref{SS:relation_U(u)} (which,
unsurprisingly, has become frequency-dependent). Again, as in the case
of standard dispersion relations---section \ref{SS:relation_U(u)}---we
can write
\begin{eqnarray}\label{A_0-I_omega}
{\cal U}_{\omega'} = {\cal U}_{H,\omega'} 
- A_0 e^{-\kappa_{\omega'} (u_{\omega}-\bar u_{I,\omega'})/c}~,
\end{eqnarray}
where $\bar u_{I,\omega'}$ is essentially $u_{I,\omega'}$ 
(with possible higher-order corrections). Assuming that the collapse
takes place rapidly, it is a good approximation to replace
$\bar u_{I,\omega'}$ by $\bar u_{I,\omega'_c}$. Actually, as we will see
shortly, this is a conservative estimate, in the sense that it
slightly underestimates the superluminal correction to the radiation
spectrum.

\subsection{Wave packet formulation}
\label{SS:wave-mod}
%-------------------------------------------------------------------
The relation ${\cal U}_{\omega'}(u_{\omega})$ we are considering interpolates
between a linear behaviour at early times and an exponential behaviour
at late times.  It has the same form as the relation $U(u)$ for
standard dispersion relations, and so we can continue following the
steps of the standard case. In particular, the equivalent of \eqref{U-u_step1} is obtained by integrating out $u_\omega$. Note that we use the subscript $\omega$ to emphasize that the $u_\omega$ are not the null coordinates of the standard case, but this should not be interpreted as an explicit function of $\omega$ and so does not complicate the integration steps in $u_\omega$ and $\omega$. However, when integrating $|\beta_{\omega_{j},u_{l};\omega'}|^2$ in $\omega'$ to obtain $N_{\omega_{j},u_{l}}$, see section~\ref{SS:wave-standard}, we must carefully consider the
frequency-dependence of the relevant terms, i.e., of ${\cal
U}_{H,\omega'}$, $\bar{u}_{I,\omega'}$ and $\kappa_{\omega'}$. The term
carrying ${\cal U}_{H,\omega'}$ is moduloed away in \eqref{U-u_step2}, and we have replaced $\bar{u}_{I,\omega'}$ by $\bar{u}_{I,\omega'_c}$, so the only relevant frequency-dependent factor that we are left with is the surface gravity $\kappa_{\omega'}$.

Moreover, because of the critical frequency $\omega'_c$ in the horizon formation
process, a finite upper boundary will also be induced in the integral. Indeed, frequencies $\omega'>\omega'_c$ do not contribute
to the radiation at all, since they do not experience a horizon. This
is a delicate but crucial point. It was already observed long ago
by Jacobson~\cite{Jacobson:1991gr} that trying to solve the
transplanckian problem naively by imposing a cut-off frequency would
seemingly extinguish Hawking radiation on a relatively short time scale. In our
case, however, this cut-off is not imposed {\it ad hoc}, but appears
explicitly because of the superluminal character of the system at high
frequencies. Moreover, the critical frequency, and hence the upper boundary
induced in the integral, depend directly on the physics inside the
horizon. Indeed, given a certain velocity profile, and in particular
its behaviour near the center of the black hole, the critical
frequency can be calculated by setting $c_k=|v|$ in eq.\eqref{group_speeds} and extracting the corresponding critical frequency from the dispersion relation~\eqref{dispersion}. We will see this effect graphically in
section~\ref{S:graphics}.

In analogy with eqs.
\eqref{U-u_step2} and \eqref{N-standard}, we now obtain the number of particles detected for each frequency $\omega_j$ as
\begin{align}\label{N-z}
N_{\omega_{j},u_{l}}&=\int_0^{\omega_c'} \d\omega'~
|\beta_{\omega_{j},u_{l};\omega'}|^2
\\
&\hspace*{-8mm}\approx {c\Delta\omega \over 2\pi}
\int_0^{\omega_c'} {\d\omega' \over \omega'}
{1 \over \kappa_{\omega'}}
{\sin^2\left[{\kappa_0 \over \kappa_{\omega'}}(z -z_{l,\omega'})\right] 
\over \left[{\kappa_0 \over \kappa_{\omega'}}(z -z_{l,\omega'})\right]^2}\;{1 \over \exp({2\pi c\omega_j
\over \kappa_{\omega'}})-1}\nonumber,
\end{align}
where now \mbox{$z= {c\Delta \omega \over 2 \kappa_0}\ln \omega' A_0$}, with
\mbox{$\kappa_0\equiv\kappa_{\omega'=0}$} (which corresponds to the standard $\kappa$ of
\eqref{standard-kappa}), and \mbox{$z_{l,\omega'}= {\kappa_{\omega'} \over \kappa_0} {\Delta \omega \over 2}(u_l-\bar u_{I,\omega'_c})$}.
Changing the integration variable from $\omega'$ to $z$, we finally obtain the central expression in our analysis:
\begin{equation}\label{N-y}
N_{\omega_{j},u_{l}}=
\frac{1}{\pi}\int_{-\infty}^{z_c} \d z 
{\kappa_0 \over \kappa_{\omega'}}
{\sin^2\left[{\kappa_0 \over \kappa_{\omega'}}(z -z_{l,\omega'})\right] 
\over \left[{\kappa_0 \over \kappa_{\omega'}}(z -z_{l,\omega'})\right]^2}\;
{1 \over \exp({2\pi c\omega_j \over \kappa_{\omega'}})-1}.
\end{equation}
Note that, as we indicated earlier, the use
of $\bar u_{I,\omega'_c}$ in the definition of $z_{l,\omega'}$ is a conservative
stance. Indeed, strictly speaking, we should write
$z_{l,\omega'}\propto(u_l-\bar u_{I,\omega'})$. For a fixed $z_c$, a smaller
value of $z_{l,\omega'}$, and hence a larger value of $\bar u_{I,\omega'}$,
means that a larger part of the central peak of the integrand will be
integrated over. Since we are replacing $\bar u_{I,\omega'}$ by
the upper bound $\bar u_{I,\omega'_c}$, we are overestimating the resulting radiation (i.e., underestimating the
modification with respect to the standard Hawking radiation).

The expression \eqref{N-y} brings out the two crucial factors mentioned earlier, and a third, corollary one.
\begin{itemize}
\item First, it shows the dependence of the total radiation on the critical frequency
$\omega'_c$ (through the integration boundary $z_c$ induced by it), as discussed just before obtaining formula \eqref{N-z}.
\item Second, it shows the importance of the frequency-dependent $\kappa_{\omega'}$
(to be compared with the fixed $\kappa$ of the standard case). Given a concrete
profile $v(t,x)$, the frequency-dependence of $\kappa_{\omega'}$ can be derived
explicitly, as we will discuss next.
\item Finally, as a corollary of the first point, it shows that, as $u_l$ (and hence $z_{l,\omega'}$) increases, a smaller part of the central peak of the integrand will be integrated over, so the radiation will die off as $u_l$ advances.
\end{itemize}

\subsection{Surface gravity}
%-------------------------------------------------------------------
A careful analysis has shown that, so far, most formulas for the
standard case could be adapted to superluminal dispersion relations by replacing the relevant magnitudes with their
frequency-dependent counterparts. For example,
$\beta_{\omega\omega'}$ was obtained by replacing $U$ and $u$ by
${\cal U}_{\omega'}$ and $u_{\omega}$, and in particular $U_H$, $u_I$
and $\kappa$ by ${\cal U}_{H,\omega'}$, $\bar{u}_{I,\omega'}$ and
$\kappa_{\omega'}$, respectively.
Given a concrete profile, we can explicitly deduce the relation
between $\kappa_{\omega'}$ and $\omega'$, with the Planck scale $k_P$
as a parameter, as follows. The horizon for a particular initial
frequency $\omega'$ is formed when 
\begin{eqnarray}
1+{k_H^2 \over k_P^2}={|v(x_{H,\omega'})|^2 \over c^2}~,
\end{eqnarray}
i.e., when $v_{ph}=c_k+v=0$, where we have used the phase velocity for
concreteness. But again, qualitatively, our results would be similar if
taking the group velocity $v_g$ instead of $v_{ph}$.

Taking into account that $k_H=k'$, the dispersion relation in the asymptotic past can be written as
\begin{eqnarray}\label{omega-v_H}
\omega'^2= 
|v(x_{H,\omega'})|^2 k_P^2 \left({|v^2(x_{H,\omega'})|^2 \over c^2}-1\right)~.
\end{eqnarray}
Given a concrete profile, $x_{H,\omega'}$ can then be obtained and $\kappa_{\omega'}$
calculated explicitly. 

For the profiles of the first type, see fig.~\ref{Fig:v-constant-kappa}, the result is trivial. Indeed, $|\bar{v}(x)|$ increases linearly between the horizon corresponding to $\omega'=0$ and the one for $\omega'_c$, so we obtain a constant $\kappa_{\omega'}=\kappa_0$ for all $\omega'<\omega'_c$.

\begin{figure}
\begin{center}
\includegraphics[width=0.75\columnwidth,height=0.5\columnwidth]{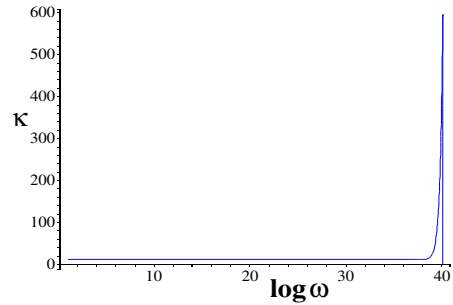}
\end{center}
\bigskip
\caption{\label{Fig:kappa-omega} Surface gravity $\kappa_{\omega'}$ with respect to the (logarithm of the) frequency for a Schwarzschild-type black hole as in fig.~\ref{Fig:v-increasing-kappa}, with the Planck scale $k_P=10^{39}~{\rm t}^{-1}$ and the critical frequency $\omega'_c=13\times 10^{39}~{\rm t}^{-1}$
(where ${\rm t}$ represents an arbitrary time unit.)}
\end{figure}

For a Schwarzschild profile as in fig.~\ref{Fig:v-increasing-kappa}, on the other hand,
\begin{eqnarray}
\bar v(x)= %-c\sqrt{2M/c^2\over x+2M/c^2}= 
-c\sqrt{c^2/2 \kappa_0 \over x+c^2/2 \kappa_0}~,
\end{eqnarray}
so we obtain
\begin{eqnarray}
\kappa_{\omega'}&\equiv &c \bigg|{d \bar v \over dx} \bigg|_{x=x_{H,\omega'}}
= \kappa_0 \left({|\bar v_H| \over c}\right)^{3/2}\nonumber\\
&=&\kappa_0 {1 \over 2\sqrt{2}}\left(
1+\sqrt{1+4\frac{\omega'^2}{c^2 k_P^2}}
\right)^{3/2}~.
\end{eqnarray}
Note that $\kappa_{\omega'}$ remains nearly constant until frequencies of the order of magnitude of the Planck scale are reached, and then starts to increase rapidly, see fig.~\ref{Fig:kappa-omega}. As we will see graphically in the next section, this can have an important qualitative influence on the radiation spectrum.

%====================================================
%====================================================

We now have all the tools necessary to compute and plot
\eqref{N-y} for different parameters of the profiles described in section~\ref{S:model}.

%--------------------------------------------------------------
\section{Graphical results and discussion}
\label{S:graphics}
\begin{figure}
\begin{center}
\includegraphics[width=0.9\columnwidth]{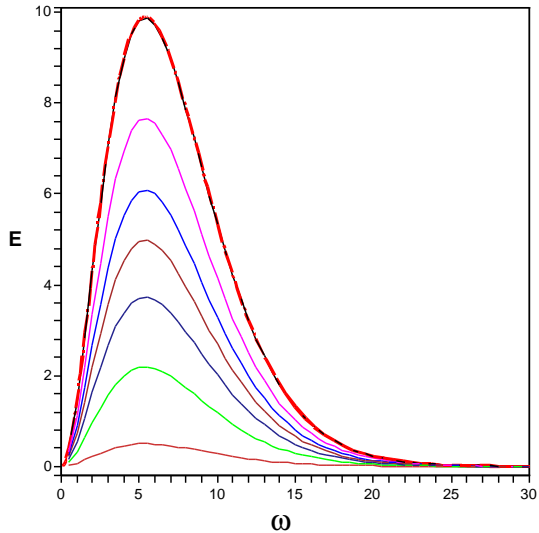}
\end{center}
\bigskip
\caption{\label{Fig:cut-off} Influence of the critical frequency on the radiation spectrum for a black hole with velocity profile such that the surface gravity is constant, as in fig.~\ref{Fig:v-constant-kappa}, and different values of the critical frequency. We have chosen $c=1$, $u_l=u_I=0$, $U_H=A=1\,{\rm t}$, $\kappa_0=12~{\rm t}^{-1}$ and the Planck scale $k_P=10^{39}~{\rm t}^{-1}$ (where ${\rm t}$ represents an arbitrary time unit), which amounts to considering a solar-mass black hole. From top to bottom we have plotted 
$\omega'_c=10^{3000}, 10^{100}, 10^{39}, 1, 10^{-39}, 10^{-100}, 10^{-300}$. Note that the standard Hawking spectrum coincides perfectly with the upper curve, which effectively corresponds to the absence of a critical frequency.}
\end{figure}

\begin{figure}
\begin{center}
\includegraphics[width=0.9\columnwidth]{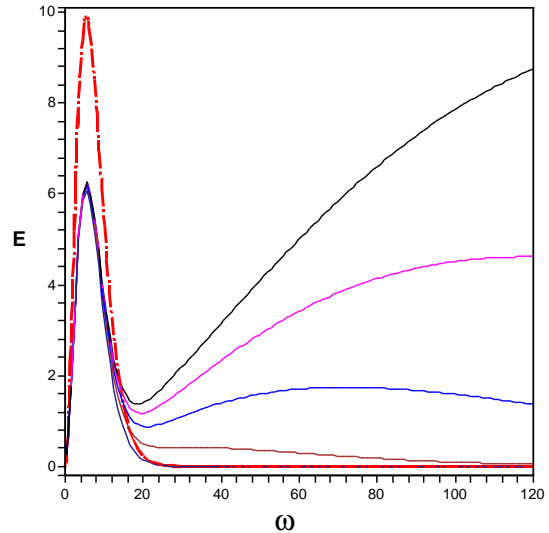}
\end{center}
\bigskip
\caption{\label{Fig:increasing-kappa} Influence of a frequency-dependent surface gravity $\kappa_{\omega'}$ on the radiation spectrum for a black hole with Schwarzschild-type velocity profile (surface gravity increases with frequency), as in fig.~\ref{Fig:v-increasing-kappa}, and different values of the critical frequency around the Planck scale $k_P$: \mbox{$\omega'_c/c k_P = 13, 10, 7, 4, 0.1$} (from top to bottom). The standard Hawking spectrum is depicted in dashed--dotted line for comparison. Numerical values of $\kappa_0$, $k_P$ etc as in fig.~\ref{Fig:cut-off}.
}
\end{figure}

\begin{figure}
\begin{center}
\includegraphics[width=0.9\columnwidth]{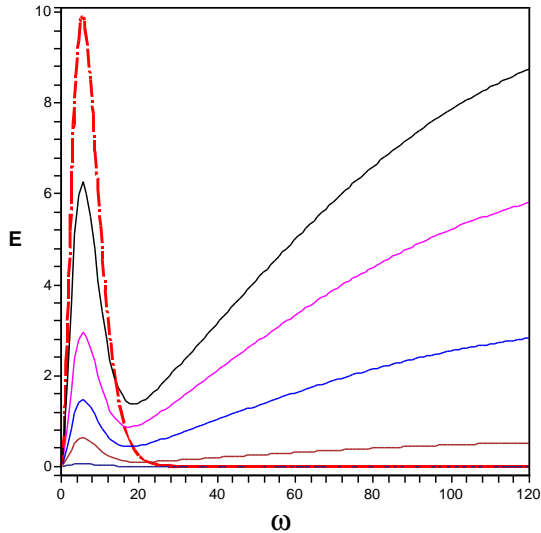}
\end{center}
\bigskip
\caption{\label{Fig:u_l} Influence of the measuring time $u_l$ on the radiation spectrum, for a Schwarzschild-type velocity profile, see fig.~\ref{Fig:v-increasing-kappa} and $\omega'_c/ck_P=13$, compare with fig.~\ref{Fig:increasing-kappa}.
Different values of $u_l$ (from top to bottom): $u_l=0,22,35,50,500~{\rm t}$
with ${\rm t}$ denoting an arbitrary time unit. The standard Hawking spectrum is depicted in dashed--dotted line for comparison. Numerical values of $\kappa_0$, $k_P$ etc as in fig.~\ref{Fig:cut-off}.
}
\end{figure}

%--------------------------------------------------------------
We have calculated eq.~\eqref{N-y} numerically using the Gauss-Chebyshev quadrature method. The results are plotted in
figs.~\ref{Fig:cut-off}--\ref{Fig:u_l} and perfectly illustrate the three important factors that we deduced theoretically in section~\ref{SS:wave-mod}. Note that in all the figures we have plotted $E\equiv\omega^3\times N$ against $\omega$ to make visual comparison with the usual thermal spectra in 3+1 dimensions easier.

Fig.~\ref{Fig:cut-off} shows the influence of the critical frequency
$\omega'_c$ for a profile with constant $\kappa_{\omega'}$, as in
fig.~\ref{Fig:v-constant-kappa}, and $u_l =\bar
u_{I,\omega_c'}$ (i.e., immediately after the horizons have formed for all $\omega'<\omega'_c$). The constant $\kappa_{\omega'}$ guarantees that the
form of the thermal spectrum is preserved. However, the intensity of
the radiation decreases with decreasing critical frequency. Actually, only for
extremely high critical frequencies is the original Hawking spectrum
recovered. For a critical frequency still well above the Planck scale, the
decrease can be significant. For example, for $\omega'_c=10^{61} c
k_P$, the peak intensity decreases by nearly 30\%, while for
$\omega'_c=c k_P$, the decrease is approximately 40\%. At the other end, note
that extremely low critical frequencies still leave a significant amount of
radiation. For example, for $\omega'_c=10^{-139} c k_P$, still 20\% of
the original peak radiation is obtained. So Hawking radiation receives significant contributions from an extremely wide range of frequencies.

%====================================================
%====================================================

Fig.~\ref{Fig:increasing-kappa} shows the radiation spectrum for a
Schwarzschild-like profile and hence increasing $\kappa_{\omega'}$, as
in fig.~\ref{Fig:v-increasing-kappa}, in particular for critical
frequencies $\omega'_c$ close to the Planck scale.  On top of the
general decrease of the standard thermal part of the spectrum
(approximately 40\%, as in the previous case) due to the finite integration boundary
induced by the critical frequency, the fact that the surface gravity
is now frequency-dependent leads to an important qualitative change of
the spectrum. The high-frequency tail of the spectrum is totally
transformed. Actually, if the critical frequency is sufficiently
higher than the Planck scale, the dominant source of radiation lies in
the high-frequency region. Note that this effect is truly a
consequence of the modification of the physics for frequencies above
the Planck scale. This can be appreciated by noticing that, at
$\omega'_c=0.1c k_P$, the whole tail-modifying effect has disappeared
and the usual thermal form of the radiation spectrum is recovered
(although still with the quantitative decrease described above).

%====================================================
%====================================================

Finally, fig.~\ref{Fig:u_l} shows the influence of the measuring time
$u_l$ (measured with respect to $\bar{u}_{I,\omega'_c}$) for a profile
of the second type (increasing $\kappa_{\omega'}$). It is clearly seen
that the radiation dies off with time, and actually dies off rather
fast. For an actual solar-mass black hole and an $\omega'_c$ of the order 
of the Planck scale, the radiation would last only a few tens of milliseconds.
Note that this effect is a corollary of the existence of a
critical frequency $\omega'_c$, since in its absence, the integration
boundaries in \eqref{N-y} would be infinite, and so the integral would
be insensitive to a change $u\to u+\Delta u$.

The combined effect of fig.~\ref{Fig:increasing-kappa} and
fig.~\ref{Fig:u_l} leads to the following qualitative picture for the
further collapse towards a singularity once the initial or classical
horizon has formed. As the interior gradually uncovers a larger
portion of the Schwarzschild geometry, two competing processes will
take place. On the one hand, the spectrum acquires ever larger
contributions associated with higher and higher temperatures. On the
other hand, the overall magnitude of the spectrum is damped with
time. The question of which process dominates would depend on the fine
details of the dynamics of the collapse, and might be further
complicated by backreaction effects, which we have not considered in
our analysis.

%====================================================
%====================================================

%--------------------------------------------------------------
\section{Summary and conclusions}
\label{S:conclusion}
%--------------------------------------------------------------
We have discussed the Hawking radiation for a collapsing configuration with
superluminal dispersion relations. Modifications of the dispersion relation
cause the horizon, and various associated quantities such as the surface
gravity $\kappa$, to become frequency-dependent. In particular, a critical
frequency $\omega'_c$ naturally appears such that frequencies higher than $\omega'_c$
do not experience a horizon at all. More generally, it also means that the standard
geometric concepts traditionally used to study black holes must be handled with
care. Nevertheless, through a detailed analysis, we have seen that the equations
related to the late-time radiation can be adapted quite straightforwardly from
standard (relativistic) to superluminal dispersion relations. 

We analytically derived an approximate equation for the particle production at
late times \eqref{N-y} with superluminal modifications of the dispersion relations. This equation clearly showed that important
modifications in the late-time radiation should be expected, first, due to the
existence of the critical frequency $\omega'_c$ and the finite upper boundary it induces in the integration, and second, due to the
frequency-dependence of $\kappa_{\omega'}$. We integrated \eqref{N-y}
numerically, and plotted the resulting spectrum, thereby confirming these expectations. 

We have seen that the standard Hawking spectrum is recovered only in a very
particular case: in the limit when the critical frequency goes to infinity (i.e.,
when the profile for the velocity $|v|$ goes to a singularity) and moreover the
surface gravity $\kappa_{\omega'}$ is constant (linear velocity profiles of the type of
fig.~\ref{Fig:v-constant-kappa}). For lower critical frequencies, as long as $\kappa$
remains constant, the thermal form of the Hawking spectrum is maintained, but
the intensity decreases rapidly with decreasing $\omega'_c$. A non-negligible
radiation persists, however, even for extremely low critical frequencies. This clearly establishes the first point, namely the importance of the critical frequency $\omega'_c$, which illustrates the statement made in the introduction that with superluminal dispersion relations, the interior of the black hole is probed, which can significantly affect the Hawking process.

For a Schwarzschild-like profile, the velocity profile generally leads to a frequency-dependent surface gravity $\kappa_{\omega'}$, increasing with the frequency $\omega'$. This means that the standard Hawking result
cannot be recovered for a Schwarzschild black hole with superluminal dispersion
relations, even when the velocity profile has a singularity. Actually, when the
velocity profile has a limiting value, the same quantitative decrease of the
Hawking part of the spectrum as before shows up. Moreover, if the critical frequency
$\omega'_c$ is above the Planck scale, a drastic qualitative change of the
radiation spectrum takes place, and for sufficiently high $\omega'_c$ (roughly a
few times the Planck scale) the high-frequency part of the spectrum even becomes
dominant. This shows the importance of the surface gravity, which again illustrates the role played by the interior of the black hole.

Finally, we have also seen that, as a corollary of the existence of a critical frequency and of the finite upper boundary induced by it in the thermal response function, the radiation spectrum dies off as time advances.

A few observations might be useful to connect our work with
existing results on Hawking radiation and its sensitivity to modified
dispersion relations. The most general observation is that the
`robustness' of Hawking radiation, which is often considered to be a
well-established result, actually depends crucially on a series of
assumptions.  These assumptions might be reasonable in the case of
subluminal dispersion relations. But for the case of superluminal
dispersion relations, as we have shown explicitly, the assumptions
needed to reproduce the standard Hawking result depend on the physics
inside the (zero-frequency) horizon, and moreover in a way which is
not compatible with the Schwarzschild geometry. In particular, in~\cite{Corley:1997pr}, it was shown for a stationary scenario that Hawking radiation is
robust with respect to superluminal modifications of the dispersion relations, provided that
positive free fall frequency modes were in their ground state just before crossing the horizon. However, it was also admitted that it is not clear whether this is the physically correct quantum state condition. In~\cite{Unruh:2004zk}, three explicit assumptions were given for the previous condition to hold: freely falling frame, ground
state and adiabatic evolution. In the same article, it is similarly pointed out that these
assumptions might fail for a superluminal dispersion relation, since (as we also mentioned in the introduction) for high-frequency modes this implies that one makes an assumption about the physics at the singularity. Rather than imposing any
conditions on the state near the horizon, and hence ultimately at the singularity, we have fixed the {\it
initial} geometry to be Minkowskian and the quantum field to be in the natural associated vacuum state, and evolved this into a black
hole configuration. Since our results are seemingly in contradiction with the
ones obtained in~\cite{Corley:1997pr} and~\cite{Unruh:2004zk}, it is worth examining in which sense the
conditions stated by those authors are violated in our approach. If one
considers a collapse scenario, for example of a BEC, in a laboratory setting,
then it is quite natural to assume that the ``freely falling frame'' condition
will be violated. Indeed, the superluminal modification
is in this case associated with the existence of a privileged external reference
frame: the black hole rest frame or lab frame (as we noted in sec.~\ref{SS:generalization_inner_product}), and not the freely falling frame, as assumed
in~\cite{Unruh:2004zk}. Note that this violation of the free-fall frame condition automatically implies a
violation of the ground state condition in the sense in which this condition is
formulated in~\cite{Unruh:2004zk}, namely as the ground state ``with respect to the freely
falling frame''. Indeed, our results suggest the following picture. The
low-energy modes experience the classical geometry, are therefore dragged along
in the free-falling frame and so at the horizon they occupy the vacuum state associated with this free-falling frame, namely the Unruh vacuum. Hence they  contribute to the black hole radiation in the
traditional Hawking way. However, the ultra-high-frequency modes (above the critical
frequency) do not see the horizon. Hence they do not couple to the classical
geometry of the collapse, but rather remain connected to the external or
laboratory frame, and therefore pass through the black hole (nearly) undisturbed. The ground state of these high-frequency modes is then not the vacuum associated with the freely falling frame, but the Boulware vacuum associated with the original Minkowski geometry, or in other words, with the stationary reference frame of the lab. So these frequencies above the critical frequency do not contribute to the thermal output spectrum. The overall radiation is a
convolution of the contributions from all the different initial frequencies,
where the surface gravity for each frequency can be interpreted as leading to an effective
weighting factor. Depending on the internal physics of the black hole, this
leads to either a spectrum of the traditional Hawking form but with a reduced
intensity (in the case of a constant surface gravity), or a modified spectrum
where the high-frequency contributions dominate (as for a Schwarzschild profile,
provided that the critical frequency lies sufficiently above the Planck scale). In any case, what our analysis shows 
with respect to the situation treated in~\cite{Corley:1997pr} and~\cite{Unruh:2004zk} is the following. The robustness of Hawking radiation discussed by those authors is the robustness with respect to a superluminal modification under the precise conditions occurring in the standard case, in which all modes occupy the ground state associated with the free-falling frame just before crossing the horizon. This assumption implies that all modes, regardless of their frequency, occupy the Unruh vacuum. Our analysis shows precisely that the superluminality can lead to a spontaneous breaking of these conditions. The natural evolution of a collapsing configuration is such that the frequencies above the critical frequency $\omega'_c$ are not in the Unruh vacuum associated with the freely falling frame, but remain in the Boulware vacuum associated with the initial asymptotic Minkowski condition. It should then come as no surprise that this can indeed have an important impact on the resulting radiation spectrum.

We should mention that
in a recent paper by Sch\"{u}tzhold and Unruh~\cite{Schutzhold:2008tx}
based on an alternative approximation method, a new potential problem
with superluminal dispersion relations is highlighted. This
``ultra-violet catastrophe'' is associated with a
higher-than-linear growth of the group velocity at large wave
numbers, and should therefore not occur for the dispersion relation
used in our work. However, we have found that a Schwarzschild interior
geometry could give place to the same sort of phenomenology.

A final observation concerns the recent
article~\cite{Carusotto:2008ep}, in which the authors numerically
simulate the formation of an acoustic horizon in a BEC, and analyze
the creation of correlated pairs of phonons through the so-called
truncated Wigner method. Our findings for the case of a constant
surface gravity seem to be in qualitative disagreement with the
discussion presented in~\cite{Carusotto:2008ep}, since the authors of
this paper assert having observed a stationary Hawking flux, while our
analysis concludes that the radiation should quickly fade off. The
source of this apparent discrepancy might reside in the fact that the
correlation function which they study is normalized, and therefore
probes the form of the spectrum (which we also find stationary with
time), but does not provide details about the net amount of particle
production.

%--------------------------------------------------------------
\section*{Acknowledgements}
%--------------------------------------------------------------

We kindly thank Stefano Liberati and Matt Visser for enlightening
comments and discussions, and Jos\'{e} Luis Jaramillo for very useful
suggestions on the numerical aspects of this work. G.J. would also
like to express his gratitude to Fernando Barbero, Daniel G\'{o}mez
Vergel and Lorenzo Sindoni for useful discussions, and SISSA (Trieste)
for hospitality. Financial support was provided by the Spanish MEC
through the projects FIS2005-05736-C03-01, FIS2005-05736-C03-02 and
FIS2006-26387-E.

%============================================================

%============================================================
%-------------------------------------------------------------

\begin{thebibliography}{99}
%============================================================
%\cite{Jacobson:2005bg}
\bibitem{Jacobson:2005bg}
  T.~Jacobson, S.~Liberati and D.~Mattingly,
  ``Lorentz violation at high energy: Concepts, phenomena and astrophysical
  constraints,''
  Annals Phys.\  {\bf 321}, 150 (2006)
  [arXiv:astro-ph/0505267].
  %%CITATION = APNYA,321,150;%%
%-------------------------------------------------------------
%\cite{Mavromatos:2007xe}
\bibitem{Mavromatos:2007xe}
  N.~E.~Mavromatos,
  ``Lorentz Invariance Violation from String Theory,''
  PoS QG-Ph:027 (2007)
  [arXiv:0708.2250 [hep-th]].
  %%CITATION = ARXIV:0708.2250;%%
%-------------------------------------------------------------
%\cite{Gambini:1998it}
\bibitem{Gambini:1998it}
  R.~Gambini and J.~Pullin,
  ``Nonstandard optics from quantum spacetime,''
  Phys.\ Rev.\  D {\bf 59}, 124021 (1999)
  [arXiv:gr-qc/9809038].
  %%CITATION = PHRVA,D59,124021;%%
%-------------------------------------------------------------
%\cite{Ashtekar:2007px}
\bibitem{Ashtekar:2007px}
  A.~Ashtekar,
  ``Loop quantum gravity: Four recent advances and a dozen frequently asked
  questions,''
  arXiv:0705.2222 [gr-qc].
  %%CITATION = ARXIV:0705.2222;%%
%-------------------------------------------------------------
%\cite{Barcelo:2005fc}
\bibitem{Barcelo:2005fc}
  C.~Barcel\'{o}, S.~Liberati and M.~Visser,
  ``Analogue gravity,''
  Living Rev.\ Rel.\  {\bf 8}, 12 (2005)
  [arXiv:gr-qc/0505065].
  %%CITATION = 00222,8,12;%%
%Living Rev. Relativity {\bf 8}, 12 (2005),
http://www.livingreviews.org/lrr-2005-12.
%-------------------------------------------------------------
%\cite{Volovik:2003fe}
\bibitem{Volovik:2003fe}
  G.~E.~Volovik,
  ``The Universe in a helium droplet,''
  Clarendon Press, Oxford (2003).
  %%CITATION = IMPHA,117,1;%%
%-------------------------------------------------------------
%\cite{Klinkhamer:2005cp}
\bibitem{Klinkhamer:2005cp}
  F.~R.~Klinkhamer and G.~E.~Volovik,
  ``Merging gauge coupling constants without grand unification,''
  Pisma Zh.\ Eksp.\ Teor.\ Fiz.\  {\bf 81}, 683 (2005)
  [JETP Lett.\  {\bf 81}, 551 (2005)]
  [arXiv:hep-ph/0505033].
  %%CITATION = JTPLA,81,551;%%
%-------------------------------------------------------------
%\cite{Mattingly:2005re}
\bibitem{Mattingly:2005re}
  D.~Mattingly,
  ``Modern tests of Lorentz invariance,''
  Living Rev.\ Rel.\  {\bf 8}, 5 (2005)
  [arXiv:gr-qc/0502097].
  %%CITATION = 00222,8,5;%%
%-------------------------------------------------------------
%\cite{Magueijo:2001cr}
\bibitem{Magueijo:2001cr}
  J.~Magueijo and L.~Smolin,
  ``Lorentz invariance with an invariant energy scale,''
  Phys.\ Rev.\ Lett.\  {\bf 88}, 190403 (2002)
  [arXiv:hep-th/0112090].
  %%CITATION = PRLTA,88,190403;%%
%-------------------------------------------------------------
%\cite{Hawking:1974rv}
\bibitem{Hawking:1974rv}
  S.~W.~Hawking,
  ``Black hole explosions,''
  Nature {\bf 248}, 30 (1974).
%-------------------------------------------------------------
%\cite{Hawking:1974sw}
\bibitem{Hawking:1974sw}
  S.~W.~Hawking,
  ``Particle Creation By Black Holes,''
  Commun.\ Math.\ Phys.\  {\bf 43} 199 (1975).
%-------------------------------------------------------------
%\cite{Agullo:2006um}
\bibitem{Agullo:2006um}
  I.~Agullo, J.~Navarro-Salas, G.~J.~Olmo and L.~Parker,
  ``Short-distance contribution to the spectrum of Hawking radiation,''
  Phys.\ Rev.\  D {\bf 76}, 044018 (2007)
  [arXiv:hep-th/0611355].
  %%CITATION = PHRVA,D76,044018;%%
%-------------------------------------------------------------
%\cite{Unruh:1980cg}
\bibitem{Unruh:1980cg}
  W.~G.~Unruh,
  ``Experimental black hole evaporation,''
  Phys.\ Rev.\ Lett.\  {\bf 46}, 1351 (1981).
%-------------------------------------------------------------
%\cite{Jacobson:1991gr}
\bibitem{Jacobson:1991gr}
  T.~Jacobson,
  ``Black hole evaporation and ultrashort distances,''
  Phys.\ Rev.\  D {\bf 44}, 1731 (1991).
%-------------------------------------------------------------
%\cite{Jacobson:1993hn}
\bibitem{Jacobson:1993hn}
  T.~Jacobson,
  ``Black hole radiation in the presence of a short distance cutoff,''
  Phys.\ Rev.\  D {\bf 48}, 728 (1993)
  [arXiv:hep-th/9303103].
  %%CITATION = PHRVA,D48,728;%%
%-------------------------------------------------------------
%\cite{Unruh:1994je}
\bibitem{Unruh:1994je}
  W.~G.~Unruh,
  ``Sonic Analog Of Black Holes And The Effects Of High Frequencies On Black
  Hole Evaporation,''
  Phys.\ Rev.\  D {\bf 51}, 2827 (1995).
%-------------------------------------------------------------
%\cite{Brout:1995wp}
\bibitem{Brout:1995wp}
  R.~Brout, S.~Massar, R.~Parentani and P.~Spindel,
  ``Hawking Radiation Without Transplanckian Frequencies,''
  Phys.\ Rev.\  D {\bf 52}, 4559 (1995).
  [arXiv:hep-th/9506121].
%-------------------------------------------------------------
%\cite{Corley:1996ar}
\bibitem{Corley:1996ar}
  S.~Corley and T.~Jacobson,
  ``Hawking Spectrum and High Frequency Dispersion,''
  Phys.\ Rev.\  D {\bf 54}, 1568 (1996).
  [arXiv:hep-th/9601073].
%-------------------------------------------------------------
%\cite{Corley:1996nw}
\bibitem{Corley:1996nw}
  S.~Corley,
  ``Particle creation via high frequency dispersion,''
  Phys.\ Rev.\  D {\bf 55}, 6155 (1997).
  %%CITATION = PHRVA,D55,6155;%%
%-------------------------------------------------------------
%\cite{Corley:1997pr}
\bibitem{Corley:1997pr}
  S.~Corley,
  ``Computing the spectrum of black hole radiation in the presence of high
  frequency dispersion: An analytical approach,''
  Phys.\ Rev.\  D {\bf 57}, 6280 (1998).
  [arXiv:hep-th/9710075].
%-------------------------------------------------------------
%\cite{Unruh:2004zk}
\bibitem{Unruh:2004zk}
  W.~G.~Unruh and R.~Sch\"{u}tzhold,
  ``On the universality of the Hawking effect,''
  Phys.\ Rev.\  D {\bf 71}, 024028 (2005).
  [arXiv:gr-qc/0408009].
%-------------------------------------------------------------
%\cite{Schutzhold:2008tx}
\bibitem{Schutzhold:2008tx}
  R.~Sch\"{u}tzhold and W.~G.~Unruh,
  ``On the origin of the particles in black hole evaporation,''
  Phys.\ Rev.\  D {\bf 78}, 041504 (2008)
  [arXiv:0804.1686 [gr-qc]].
  %%CITATION = PHRVA,D78,041504;%%
%-------------------------------------------------------------
%\cite{Visser:2001kq}
\bibitem{Visser:2001kq}
  M.~Visser,
  ``Essential and inessential features of Hawking radiation,''
  Int.\ J.\ Mod.\ Phys.\  D {\bf 12}, 649 (2003).
  [arXiv:hep-th/0106111].
%-------------------------------------------------------------
%\cite{Barcelo:2006yi}
\bibitem{Barcelo:2006yi}
  C.~Barcel\'{o}, A.~Cano, L.~J.~Garay and G.~Jannes,
  ``Stability analysis of sonic horizons in Bose-Einstein condensates,''
  Phys.\ Rev.\  D {\bf 74}, 024008 (2006)
  [arXiv:gr-qc/0603089].
  %%CITATION = PHRVA,D74,024008;%%
%-------------------------------------------------------------
%\cite{Corley:1998rk}
\bibitem{Corley:1998rk}
  S.~Corley and T.~Jacobson,
  ``Black hole lasers,''
  Phys.\ Rev.\  D {\bf 59}, 124011 (1999)
  [arXiv:hep-th/9806203].
  %%CITATION = PHRVA,D59,124011;%%
%-------------------------------------------------------------
%\cite{Garay:1999sk}
\bibitem{Garay:1999sk}
  L.~J.~Garay, J.~R.~Anglin, J.~I.~Cirac and P.~Zoller,
  ``Black holes in Bose-Einstein condensates,''
  Phys.\ Rev.\ Lett.\  {\bf 85}, 4643 (2000)
  [arXiv:gr-qc/0002015].
  %%CITATION = PRLTA,85,4643;%%
%-------------------------------------------------------------
%\cite{Garay:2000jj}
\bibitem{Garay:2000jj}
  L.~J.~Garay, J.~R.~Anglin, J.~I.~Cirac and P.~Zoller,
  ``Sonic black holes in dilute Bose-Einstein condensates,''
  Phys.\ Rev.\  A {\bf 63}, 023611 (2001)
  [arXiv:gr-qc/0005131].
  %%CITATION = PHRVA,A63,023611;%%}. 
%-------------------------------------------------------------
%\cite{Barcelo:2000tg}
\bibitem{Barcelo:2000tg}
  C.~Barcel\'{o}, S.~Liberati and M.~Visser,
  ``Analog gravity from Bose-Einstein condensates,''
  Class.\ Quant.\ Grav.\  {\bf 18}, 1137 (2001)
  [arXiv:gr-qc/0011026].
  %%CITATION = CQGRD,18,1137;%%
%-------------------------------------------------------------
%\cite{Barcelo:2001ca}
\bibitem{Barcelo:2001ca}
  C.~Barcel\'{o}, S.~Liberati and M.~Visser,
  ``Towards the observation of Hawking radiation in Bose-Einstein
  condensates,''
  Int.\ J.\ Mod.\ Phys.\  A {\bf 18}, 3735 (2003)
  [arXiv:gr-qc/0110036].
  %%CITATION = IMPAE,A18,3735;%%
%-------------------------------------------------------------
\bibitem{Schutzhold:2006}
  R.~Sch\"{u}tzhold,
  ``On the detectability of quantum radiation in Bose-Einstein Condensates,''
  Phys.\ Rev.\ Lett.\ {\bf 97}, 190405 (2006)
 [arXiv:quant-ph/0602180].
%-------------------------------------------------------------
%\cite{Wuster:2007nf}
\bibitem{Wuster:2007nf}
  S.~W\"{u}ster and C.~M.~Savage,
  ``Limits to the analogue Hawking temperature in a Bose-Einstein condensate,''
  Phys.\ Rev.\  A {\bf 76}, 013608 (2007)
 [arXiv:cond-mat/0702045].
  %%CITATION = COND-MAT/0702045;%%
%-------------------------------------------------------------
%\cite{Balbinot:2007de}
\bibitem{Balbinot:2007de}
  R.~Balbinot, A.~Fabbri and S.~Fagnocchi,
  ``Non-local density correlations as signal of Hawking radiation in BEC
  acoustic black holes,''
  Phys.\ Rev.\ A {\bf 78}, 021603 (2008)
  [arXiv:0711.4520 [cond-mat.other]].
  %%CITATION = ARXIV:0711.4520;%%
%-------------------------------------------------------------
%\cite{Carusotto:2008ep}
\bibitem{Carusotto:2008ep}
  I.~Carusotto, S.~Fagnocchi, A.~Recati, R.~Balbinot and A.~Fabbri,
  ``Numerical observation of Hawking radiation from acoustic black holes in
  atomic BECs,''
  New J.\ Phys.\  {\bf 10}, 103001 (2008)
  [arXiv:0803.0507 [cond-mat.other]].
  %%CITATION = NJOPF,10,103001;%%
%-------------------------------------------------------------
%\cite{Wuester:2008kh}
\bibitem{Wuester:2008kh}
  S.~W\"{u}ster,
  ``Phonon background versus analogue Hawking radiation in Bose-Einstein
  condensates,''
 Phys.\ Rev.\ A {\bf 78}, 021601(R) (2008)
  [arXiv:0805.1358 [cond-mat.other]].
  %%CITATION = ARXIV:0805.1358;%%
%-------------------------------------------------------------
%\cite{Visser:1997ux}
\bibitem{Visser:1997ux}
  M.~Visser,
  ``Acoustic black holes: Horizons, ergospheres, and Hawking radiation,''
  Class.\ Quant.\ Grav.\  {\bf 15}, 1767 (1998)
  [arXiv:gr-qc/9712010].
  %%CITATION = CQGRD,15,1767;%%
%-------------------------------------------------------------
%\cite{Fabbri:2005mw}
\bibitem{Fabbri:2005mw}
  A.~Fabbri and J.~Navarro-Salas,
  ``Modeling black hole evaporation,''
%\href{http://www.slac.stanford.edu/spires/find/hep/www?irn=6315810}{SPIRES entry}
 Imperial College Press, London (2005).
%-------------------------------------------------------------
%\cite{Visser:2007du}
\bibitem{Visser:2007du}
  M.~Visser and S.~Weinfurtner,
  ``Analogue spacetimes: Toy models for ``quantum gravity'',''
  PoS QG-Ph:042 (2007)
  [arXiv:0712.0427 [gr-qc]].
  %%CITATION = ARXIV:0712.0427;%%
%-------------------------------------------------------------
%\cite{Visser:2007nx}
\bibitem{Visser:2007nx}
  M.~Visser,
  ``Emergent rainbow spacetimes: Two pedagogical examples,''
  arXiv:0712.0810 [gr-qc].
  %%CITATION = ARXIV:0712.0810;%%
%-------------------------------------------------------------
%\cite{Barcelo:2006np}
\bibitem{Barcelo:2006np}
  C.~Barcel\'{o}, S.~Liberati, S.~Sonego and M.~Visser,
  ``Quasi-particle creation by analogue black holes,''
  Class.\ Quant.\ Grav.\  {\bf 23}, 5341 (2006).
  [arXiv:gr-qc/0604058].
%-------------------------------------------------------------
%\cite{Barcelo:2006uw}
\bibitem{Barcelo:2006uw}
  C.~Barcel\'{o}, S.~Liberati, S.~Sonego and M.~Visser,
  ``Hawking-like radiation does not require a trapped region,''
  Phys.\ Rev.\ Lett.\  {\bf 97}, 171301 (2006).
  [arXiv:gr-qc/0607008].
%-------------------------------------------------------------
%============================================================
\end{thebibliography}
\end{document}